**REVIEW ARTICLE**            **OPEN ACCESS**

# WisdomNet: Prognosis of COVID-19 with Slender Prospect of False Negative Cases and Vaticinating the Probability of Maturation to ARDS using Posteroanterior Chest X-Rays

**Peeyush Kumar**, **Ayushe Gangal** and **Sunita Kumari***

G.B. Pant Government Engineering College, Delhi - 110 020, India.

## Abstract

Coronavirus is a large virus family consisting of diverse viruses, some of which disseminate among mammals and others cause sickness among humans. COVID-19 is highly contagious and is rapidly spreading, rendering its early diagnosis of preeminent status. Researchers, medical specialists and organizations all over the globe have been working tirelessly to combat this virus and help in its containment. In this paper, a novel neural network called WisdomNet has been proposed, for the diagnosis of COVID-19 using chest X-rays. The WisdomNet uses the concept of 'Wisdom of Crowds' as its founding idea. It is a two-layered convolutional Neural Network (CNN), which takes chest x-ray images as input. Both layers of the proposed neural network consist of a number of neural networks each. The dataset used for this study consists of chest x-ray images of COVID-19 positive patients, compiled and shared by Dr. Cohen on GitHub, and the chest x-ray images of healthy lungs and lungs affected by viral and bacterial pneumonia were obtained from Kaggle. The network not only pinpoints the presence of COVID-19, but also gives the probability of the disease maturing into Acute Respiratory Distress Syndrome (ARDS). Thus, predicting the progression of the disease in the COVID-19 positive patients. The network also slender the occurrences of false negative cases by employing a high threshold value, thus aids in curbing the spread of the disease and gives an accuracy of 100% for successfully predicting COVID-19 among the chest x-rays of patients affected with COVID-19, bacterial and viral pneumonia.

## INTRODUCTION

The newly found SARS-COV-2 virus is zoonotic in nature and causes COVID-19, as named by the WHO on 11th February 2020[1]. It has affected, as of this writing, 27,09,483 people and has claimed the lives of 1,90,872 people across the globe[4]. Though, its mortality rate is 3.4%[5] which is lower than that of SARS and MERS with 9.6% and 34.3% respectively, but it has augmented into a vicious malady because of its high reproductive number $R_0$. COVID-19 has $R_0$ of (2 - 3.5), which is much higher than SARS and MERS[6]. Testing has been called as our form of intelligence, which will give us an edge over the newly spreading coronavirus[7]. The present gold-standard for testing is the reverse transcription-polymerase chain reaction (RT-PCR), which works by detecting viral nucleic acid[8]. Unfortunately, this method of testing has been deemed faulty by many doctors and researchers and is said to have a false negative rate of up to 30%[9], which can prove to be extremely precarious especially considering the diseases's high reproductive number. Therefore, there is a dire need to come up with newer and better approaches for testing. Amid the world's constant endeavor over this contagion, it is better to be safe than sorry, therefore, a false positive can still be overlooked and made up for, but a false negative can hugely spoil all results obtained till now from the world wide lockdown. Therefore, an accurate testing method is what will lead us to win this war against COVID-19.

Deep learning has proved to be an asset and has shown potential in detecting COVID-19 using chest X-rays and computed tomography (CT)[10-15,19-22]. The approaches used so far by researchers have been able to classify chest x-rays and CTs but up to a certain level of accuracy and have employed unnecessarily heavy networks which have intense hardware requirements[10-14]. Some of the approaches are not found to be target-specific enough and were detecting pneumonia using a pneumonia specific dataset[13], while some detected COVID-19, SARS, MERS and ARDS[3] by placing them all in a single class[11], overlooking the fact that majority of the COVID-19 cases do not achieve the ferocity of SARS and MERS at the initial stage, and also that (acute respiratory distress syndrome) ARDS is a stage which is achieved in critical stages of respiratory diseases and is not an independent disease caused by a pathogen in itself[15]. But a major drawback of the researches hitherto is that they can't be used solely as a method of testing for coronavirus and requires the RT-PCR test for confirmation, thus defeating the whole purpose of presenting an alternative testing methodology[10-14].

In this paper, a novel and more reliable approach for detecting the presence of COVID-19, using chest X-rays, has been presented. A neoteric neural network called WisdomNet has been proposed, which not only detects the presence of the disease but also informs about the progression of the diseases by giving the patient's probability of approaching the ARDS stage using the chest x-ray images. The aim of this network is to minimize the occurrences of false negative cases, thus introducing a streak of surety in the testing outcomes, which by far has not been performed successfully. The founding idea of this paper is based on a concept called 'Wisdom of Crowds', which was introduced in 2004 by James Surowiecki[2]. It has been repeatedly proven that the intelligence of a cluster of people is higher than that of any single person, be it exit polls for elections, stocks and mutual funds[16], forecasting[17], businesses governing the economy[18] and so on. The WisdomNet is a two-layered neural network whose each layer incorporates 80 neural networks each. The first layer classifies the chest x-rays into COVID-19 positive or negative, and the second layer predicts the probability of maturation of the disease into ARDS for the x-rays which were previously classified as COVID-19 positive. The network successfully implements the concept of wisdom of crowds by considering the aggregate of the probabilities generated by each of the 80 networks in each layer. The results obtained are in agreement with the theory presented in the paper and an accuracy of 100% is achieved for the task of predicting COVID-19.

The paper is divided into 5 sections. Section 2 talks about the research hitherto for COVID-19 and the related concepts so as to juxtapose the previous works and the proposed system accurately. Section 3 delineates the proposed methodology with explanatory diagrams. Section 4 gives the results analysis and section 5 talks about the cogent conclusions that can be drawn from this study.

## RELATED WORKS

Authors[10] fine-tuned the top layer of famous Deep Learning architectures and compared their results using chest x-ray and CT dataset. The VGG16, VGG19, Xception, Resnet50, DenseNet201, Inception_V3, Inception_ResNet_V2 and MobileNet_V2 architectures are considered for this study. The proposed CNN architecture consisted of the input layer, convolutional layers, pooling layers, ReLu layers and the fully connected layers. The dataset used consisted of 5856 chest x-ray and CT images. Resnet50, Inception_ReseNet_V2 and MobileNet_V2 obtained the highest accuracy of 96%.

Authors[11] suggested a new methodology for detecting the presence of COVID-19 using eleven pre-trained convolutional networks. Deep features from the fully connected layer of AlexNet, VGG16, VGG19, Xception, Resnet18, Resnet50, Resnet101, Inceptionv3, Inceptionresnetv2, GoogleNet and Densenet201 were extracted and fed into Support Vector Machine (SVM) classifier for training. Combined datasets of SARS, MERS, ARDS and COVID-19 from GitHub, Kaggle and Open-i repositories were used for this study. ResNet50 plus SVM gave the best results among all networks.

Authors[12] used Generative Adversarial Network (GAN) along with deep transfer learning networks like AlexNet, GoogleNet, Resnet18 and Squeezenet for classifying COVID-19 on a pneumonia dataset containing 5863 chest x-ray images. The images for training were generated by the GAN using the original dataset, and these images were used to train the deep transfer networks. The Resnet18 achieved the highest accuracy of 99%.

Authors[13] predicted the presence of viral pneumonia using 5863 chest x-ray images, assuming that a patient with viral pneumonia during the time of the epidemic has a high probability of being COVID-19 positive. Pre-trained CNN architectures like ResNet50, ResNet34, VGG19, InceptionResNetV2 and DenseNet169 were used as classifiers. A bidirectional Long-Short Term Memory (LSTM-RNN) was used for detecting the final split images. All models were found to be at least 84% accurate and the InceptionResNetV2 detected the least number of false negative cases of viral pneumonia.

Authors[14] proposed a deep learning based framework for detecting COVID-19 using 50 chest x-ray images available on GitHub. The proposed framework made use of seven deep convolutional neural networks (DCNN) like VGG19, MobileNetV2, DenseNet201, InceptionResNetV2, InceptionV3, Xception and ResNetV2. The DenseNet201 and VGG19 obtained the highest accuracy of 90% among all.

Authors[19] proposed the use of modified CNN and a modified pre-trained AlexNet model for the detection of COVID-19 using the chest x-ray and CT images. The modified CNN consisted of a single convolutional layer with 16 filters, a batch normalization layer, ReLu function layer, two fully connected layers, softmax function layer and a layer for classification. The data used for this study was collected from 5 different sources and consisted of 170 x-ray and 361 CT images in total. The modified CNN obtained 94.1% accuracy for CT images and modified AlexNet obtained 98% accuracy for x-rays.

Authors[20] proposed a capsule networks based framework called COVID-CAPS for the detection of COVID-19 using a small dataset containing x-ray images of normal, non-COVID-19 viral, bacterial and COVID-19. The proposed framework consisted of four convolutional layers and three capsule layers. COVID-CAPS was able to achieve 95.7% accuracy.

Authors[21] presented a method based on artificial neural networks (ANN) for detecting early stages of COVID-19 by means of chest x-ray images. CNN, SVM and random forest were applied as well on the same dataset. The augmented dataset used for this study consisted of chest x-ray images for COVID-19, SARS, MERS and ARDS, and was obtained from GitHub. The accuracy obtained by CNN, SVM and random forest models were 95.2%, 90.5% and 81% respectively.

Authors[22] used a pre-trained eighteen layer residual CNN for the detection of COVID-19 using chest x-ray images. They were also able to facilitate anomaly detection for minimizing faults while classifying using an anomaly score. The model was able to obtain 96% accuracy for COVID-19 cases and 70.65% for non-COVID-19 cases.

## MATERIALS AND METHODS
### WisdomNet

The proposed WisdomNet consists of two connected layers. The first layer predicts the presence of COVID-19 in the chest x-rays and the second layer gives the probability of progression into the ARDS stage for the COVID-19 positive cases. Each layer incorporates numerous convolutional neural networks having similar architecture (see 'Architecture of convolutional Neural Network'). The founding theory of WisdomNet is the concept of Wisdom of Crowd[2]. The use of this statistical concept increases the authenticity of the predictions made by the network, and also aids in minimizing the occurrences of the false negative classifications, thus enhancing the correctness of the results obtained to a greater extent. The architecture of WisdomNet is given in fig. 1.

### Architecture of convolutional Neural Network

The convolutional neural network $N$ used here accepts input image $I$ of dimension *256x256x3*. Let each layer of convolutional used by $N$ be represented as , where $k$ is the size of the kernel used to compute feature maps and $n$ is the total number of nodes present in the layer or the total number of feature maps to be computed using the kernel of size . Also, let each layer of max pooling used be , where k is the size of the kernel used, and each dense layer or layer consisting of multiple perceptrons be , where n is the total number of perceptrons present in the layer. Therefore, the architecture of the convolutional part of the network and the output provided by it can be represented as:

$$Conv_3^{64}(Max_2(Conv_3^{64}(Max_2(Conv_7^{64}(I))))) = C_o$$

...(1)

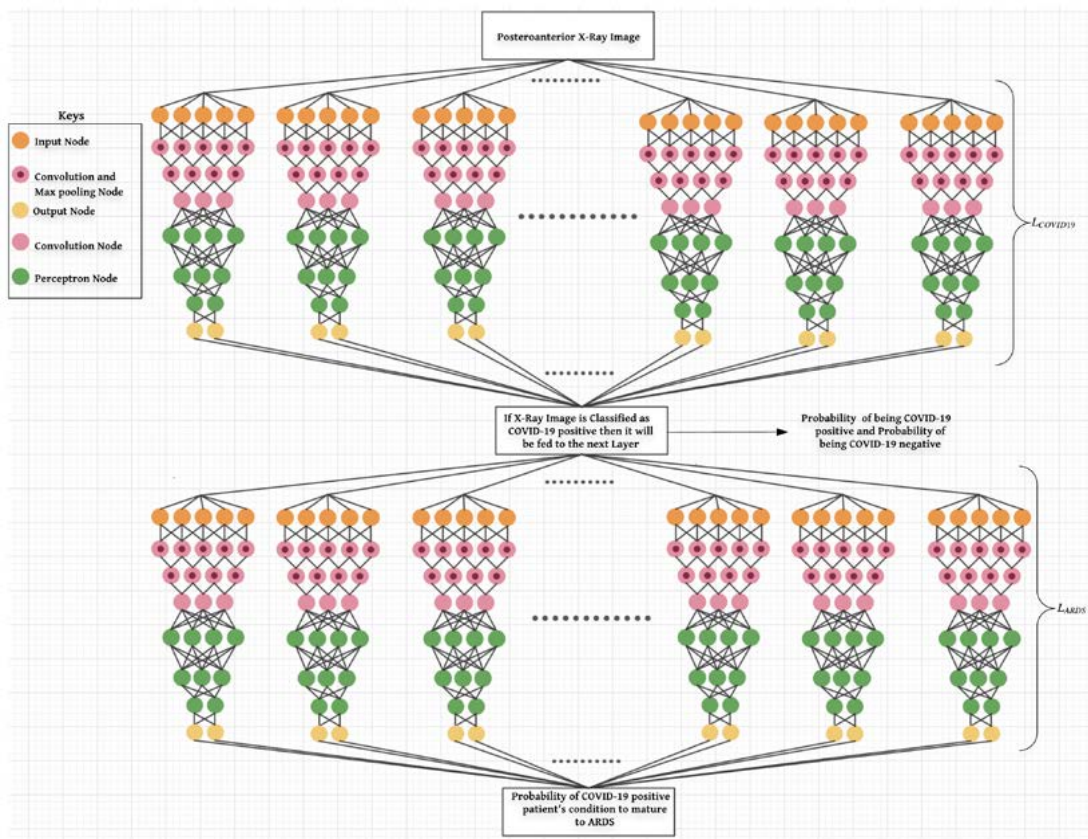

**Fig. 1.** Shows the architecture of WisdomNet having 80 convolutional neural networks in each layer

The output obtained was flattened before feeding in into the dense layer network and the final output was obtained. This can be represented using equation 2 and 3:

$$C_o \to (Flattened) \to C_o^{flattened} \quad ...(2)$$

$$D_2(D_{32}(D_{64}(C_o^{flattened}))) = P_{output} \quad ..(3)$$

Each convolutional layer in equation 1 uses ReLu[23] activation function and each dense layer in equation 2 also uses Relu except , which uses the softmax[23] activation function.

### The COVID-19 Layer

The COVID-19 layer consists of numerous convolutional neural networks having similar architecture. The layer consists of neural networks of architecture defined in the previous section. Therefore, the layer can be represented as (each neural net *N* below is represented as , where is *i* is the just the number allotted to them for representation):

$$L_{COVID19} = \{N_1, N_2, N_3, ...., N_\lambda\} \quad ...(4)$$

Each neural net *N* outputs a probability set (equation 3), which can be represented as:

$$P_{output} = [P_{Positive}, P_{Negative}] \quad ...(5)$$

Where, is the probability of being COVID-19 negative and is the probability of being COVID-19 positive.

Therefore, the probability set given by the as a whole is the mean of all probabilities given by each neural network used in this layer and can be represented as:

$$P_{output}^{Mean} = \frac{\sum_{i=1}^{\lambda} P_{output}^i}{\lambda} \quad ...(6)$$

Where, is the output provided by neural network (refer equation 4).

The threshold has been set to 50% to classify a posteroanterior x-rays as COVID-19 positive, but it has been set to 70% to classify the x-rays as COVID-19 negative. Thus, if and only if the model is more than 70% sure that a case is negative, then only it will be classified as a COVID-19 negative case, otherwise it will be classified as a positive case. By doing this, the possibility of occurrence of false negative cases minimizes dramatically. The value of was set to 80 in the experiment as it produced the best results at this value.

### The ARDS Layer

The ARDS layer also consists of a number of convolutional networks having similar architecture. Same as layer, this layer also has neural networks, having architecture defined in the 'Architecture of convolutional Neural Network' section. Therefore, the layer can also be represented as represented in equation 4. In , each neural net will output a probability set (from equation 3), which can be represented as:

$$P_{output} = [P_{Non-ARDS}, P_{ARDS}] \quad ...(7)$$

Where, is the probability that the patient has not yet reached ARDS stage and is the probability that the patient has reached ARDS stage. Higher is the value of , than higher is the patient's progression towards the ARDS stage.

Only the x-rays previously classified as COVID-19 positive by the are fed to this layer to vaticinate the probability of the condition to mature to the ARDS stage. The probability set given by the complete can also be represented by equation 6, where each singular output set is in the form given in equation 7. Here also the value of has been set to 80.

### Training of $L_{COVID-19}$

The layer is trained on a dataset consisting of posteroanterior lung x-rays, which have been affected by SARS-COV-2 virus, and has been compiled and shared by Dr. Joseph Cohen of University of Montreal on GitHub[24]. The x-ray images of non-COVID-19 lungs is obtained from Kaggle[25], out of which 30% are of patients affected with bacterial pneumonia, 30% are of patients affected with viral pneumonia and 40% the x-ray images are of healthy people. This unique distribution of data has been established in order to broaden the generalization power of the model. A total of 200 neural networks having the same architecture (defined in 'Architecture of convolutional Neural Network' section) are trained using this dataset. A sample of 80 such networks are selected out of 200 networks to form the layer. Each convolutional neural net uses Adam optimizer[26] with learning rate and the decay rate clipped to . Each neural network is trained for 4 to 10 iterations and the images of the training dataset are also augmented before feeding, in order to improve the generalization of the network. The labels of the whole dataset are one-hot-encoded[27] and each pixel in the image is mapped between 0 to 1. The Binary-Cross Entropy loss function[28] *BCE(.)* is used, and can be represented as:

$$BCE(y, P(y)) = \frac{-1}{\varphi} \sum_{i=1}^{\varphi} y_i . \log(P(y_i)) + (1 - y_i) . \log(1 - P(y_i))$$
...(8)

Where, y is the label (0 for COVID19 positive and 1 for COVID19 negative) and is the probability predicted by the network for all points .

**Training of $L_{ARDS}$**

The layer is trained on a dataset consisting of posteroanterior lung x-rays of patients whose condition had matured to the ARDS stage, which has also been compiled and shared by Dr. Joseph Cohen on GitHub. The lung x-rays images of patients whose condition hadn't matured to ARDS have also been obtained from Kaggle. Each neural net is trained for 10-15 iterations, and the rest of the training conditions are the same as the conditions given in the previous section. The loss function used for this layer can be represented using equation 8 with a slight variation in the label y (0 for non-ARDS lungs and 1 for ARDS lungs). Few x-rays of patients with mild cases of COVID-19 (whose condition hadn't progressed to the ARDS stage) are also included in the non-ARDS class in order to enhance the quality of predictions. The rationale behind training this layer on the ARDS dataset is to be able to visualize the widespread inflammation in the chest x-rays (can be seen in fig. 2), that develops in patients with acute respiratory distress syndrome (ARDS). By doing so, this layer is being trained to check the severity of inflammation in the chest x-rays of COVID-19 positive cases (given by to this layer) and therefore, predict the patients' possibility of maturation to ARDS.

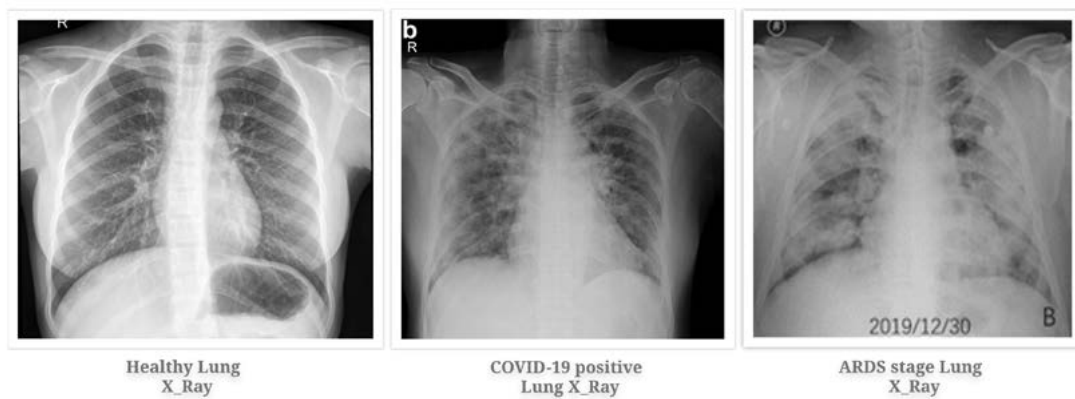

Healthy Lung X_Ray     COVID-19 positive Lung X_Ray     ARDS stage Lung X_Ray

**Fig. 2.** Shows the contrast between the inflammation of healthy, COVID-19 and ARDS stage lung x-rays[24, 25]. It can be seen that ARDS lung x-ray has more whitespots or infiltrates than COVID-19 and health lung x-ray. Thus, a neural network can predict the patient's stage by processing its x-ray image

**RESULTS AND DISCUSSION**

The utilization of the concept of Wisdom of Crowd for the prognosis of COVID-19 yields noticeably promising results. The proposed WisdomNet is successful in curbing the unnerving problem of false negative results. It also predicts the chances of advancement in the severity of the patient's condition in the future by predicting the probability of the patient's disease progressing into the ARDS stage. The results obtained by diagnosing x-rays of 10 patients using WisdomNet are shown in fig. 3. The results predicted by WisdomNet are shown in fig. 3 in detail. An enthralling thing to notice in the results shown in fig 3, is that initially Subject 4 was not a COVID-19 positive patient but as the probability predicted by the layer of WisdomNet for this case is 0.642 or 64.2% and the required probability for classifying an x-ray as COVID-19 negative has been set to 0.70 or 70%, subject 4 has been diagnosed as COVID-19

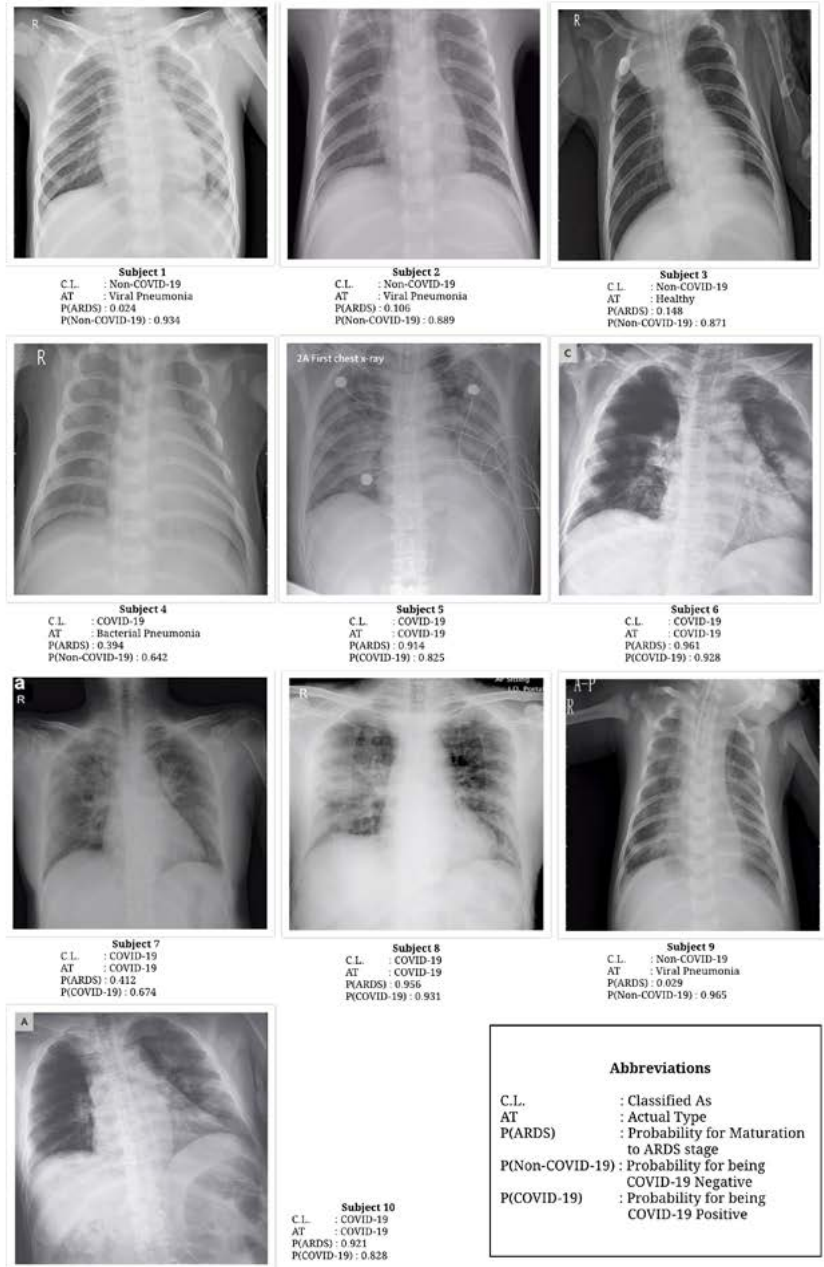

**Fig. 3.** Results obtained by diagnosing x-rays of 10 subjects[24, 25] using WisdomNet. The calculated probabilities and their respective values, with complete abbreviations are cogently given in the figure itself

positive. This scenario occured due to a logical statistical reason.

In fig. 4, graph *C* shows the normal distribution[29] curve of all the probabilities for being COVID-19 negative which have been predicted by each convolutional neural network in for subject 4, and graph *D* shows the normal distribution curve of all the probabilities for being COVID-19 positive which are again predicted by each convolutional neural network in for subject 4. It can be seen lucidly in graph *C* that the majority of probabilities predicted are less than the mean probability calculated. Also in graph *D*, the majority of probabilities predicted are greater than the mean probability calculated. Both graphs *C* and *D* show statistical scepticism when it comes to diagnosing subject 4. Such uncertainty in diagnosis increases the chances of a wrong classification, and furthermore classifying a COVID-19 positive patient as negative is even worse. So as to tackle such fluctuant eventuality, the threshold for classifying COVID-19 negative cases has been set to 70% and the astute results of such can be seen in case of subject 4.

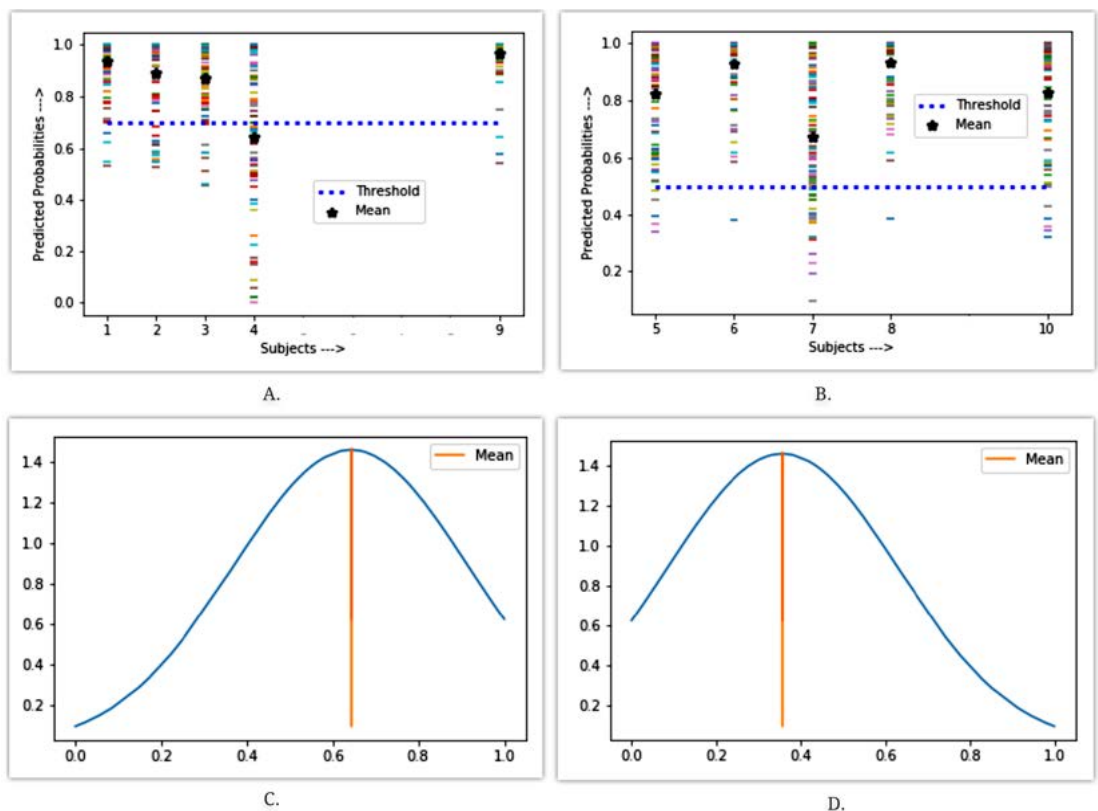

**Fig. 4.** Graph A and B shows the plot of all the probabilities predicted by 80 convolutional neural networks in , for COVID-19 negative and positive cases respectively. Graphs C and D shows the normal distribution curve of probabilities predicted for subject 4

In fig. 4, graphs *A* and *B* shows the plot of probabilities predicted by each neural network in the layer for COVID-19 negative and COVID-19 positive subjects respectively. In graph A, it can again be seen that there is high variance in probabilities predicted for subject 4 and the mean probability lies below the threshold which was conducive to it being classified as COVID-19 positive. Probabilities of maturation to the ARDS stage are also shown in fig. 3, and the authenticity of which can be empirically proved by looking at the x-ray of the same. The WisdomNet is trained on different training set proportions in order to showcase its generalization capability. The results of the same are portrayed in table 1.

**Table 1.** Performance of WisdomNet on different proportions of dataset. For example, in case number 1, the training set is 20% of the dataset and the test set is 80% of the dataset

| Case No. | Proportion of Training Set | Proportion of Test Set | Accuracy |
| --- | --- | --- | --- |
| 1 | 20% | 80% | 95% |
| 2 | 30% | 70% | 96% |
| 3 | 40% | 60% | 100% |
| 4 | 50% | 50% | 100% |
| 5 | 60% | 40% | 100% |
| 6 | 70% | 30% | 100% |
| 7 | 80% | 20% | 100% |

**CONCLUSION**

In this paper, a more dependable novel approach for the purpose of detecting the presence of COVID-19 using chest x-rays has been proposed. A novel two-layered neural network called the WisdomNet has been proposed by means of this paper, which not only classifies the chest x-rays as COVID-19 positive or negative, but also gives the probability of the disease's maturation into the ARDS stage. All patients suffering from ARDS will eventually require ventilator support as even 100% oxygen therapy doesn't seem to work[30]. The proposed model predicts the COVID-19 positive patients requiring ventilator support in advance, by predicting the progression of the disease to the ARDS stage, as higher are the chances of maturation to the ARDS stage, higher are the chances that the patient will require ventilator support.

The results obtained showcase the successful exploitation of the concept of Wisdom of Crowds for the task of predicting the presence and extent of COVID-19 in the chest x-rays. The two layers of the WisdomNet consisting of 80 neural networks each, worked in harmony to further produce more accurate results and prevent misclassification to a larger extent. Threshold value of 70% has been set for the classification of chest x-rays as COVID-19 negative in order to avoid the occurrences of false negative cases. The proposed methodology produced better and more reliable results without any heavy hardware requirements, and was shown to be target specific. The layer of the WisdomNet was admirably able to distinguish bacterial and viral pneumonia as non-COVID-19 penumonia, which further deems the proposed methodology perfect for testing purposes by medical practitioners.

The generalization power of the model is clearly depicted in table 1, and with possible future additions to the available chest x-ray datasets for COVID-19, the generalization power of the model will likely flourish even more. The presented work outputs superior results than researches done hitherto in terms of testing accuracy. It also produces statistically more accurate results than RT-PCR kits, which is a default for testing so far. Another denouement that can be drawn is that, the collective decision of many small sized neural networks using the proven theory of wisdom of crowd is found to be more authentic and accurate than a large sized neural network which requires unnecessary large and heavy hardware for computation. Therefore, the proposed system has the potential to successfully contribute in the world's crusade against COVID-19.

**ACKNOWLEDGMENTS**

None.

**CONFLICT OF INTEREST**

The listed author(s) declare no conflict of interest in any capacity, including competing or financial.

**AUTHORS' CONTRIBUTION**

All listed author(s) have made a substantial, direct and intellectual contribution to the work, and approved it for publication.

**FUNDING**

None.

**ETHICS STATEMENT**

This article does not contain any studies with human participants or animals performed by any of the authors.

**AVAILABILITY OF DATA**

Not applicable.

**REFERENCES**
1. Rolling updates on CoronaVirus Disease - updated 20th April (Events as they happen). https://www.who.int/emergencies/diseases/novel-coronavirus-2019/events-as-they-happen, Accessed 22 April 2020.
2. Surowiecki, James. 2005. The Wisdom of Crowds. New York: Anchor Books.
3. Acute Respiratory Distress Syndrome (ARDS) - Symptoms and Causes. https://www.mayoclinic.org/diseases-conditions/ards/symptoms-causes/syc-20355576, Accessed 22 April 2020
4. Coronavirus Disease Pandemic. https://www.who.int/emergencies/diseases/novel-coronavirus-2019, Accessed 22 April 2020.
5. Wang Y, Wang Y, Chen Y, Qin Q. Unique epidemiological and clinical features of the emerging 2019 novel coronavirus pneumonia (COVID-19) implicate special control measures. *Journal of Medical Virology*. 2020;92(6):568-76.
6. Liu Y, Gayle AA, Wilder-Smith A, Rocklov J. The reproductive number of COVID-19 is higher compared to SARS coronavirus. *Journal of Travel Medicine*. 2020 Mar 1.

7. Instead of just flattening the COVID-19 curve, can we 'crush' it? https://www.livescience.com/can-covid-19-be-crushed.html, Accessed 22 April 2020.
8. Ai T, Yang Z, Hou H, et al. Correlation of chest CT and RT-PCR testing in coronavirus disease 2019 (COVID-19) in China: a report of 1014 cases. *Radiology*. 2020;26:200642.
9. Coronavirus Diseases False Negatives in RT-PCR Tests, https://www.livescience.com/covid19-coronavirus-tests-false-negatives.html, Accessed 23 April 2020.
10. Asnaoui KE, Chawki Y, Idri A. Automated Methods for Detection and Classification Pneumonia based on X-Ray Images Using Deep Learning. *arXiv preprint arXiv*. 2020;2003.14363.
11. Sethy PK, Behera SK. Detection of coronavirus Disease (COVID-19) based on Deep Features. *Preprints*. 2020;2020030300:2020.
12. Khalifa NE, Taha MH, Hassanien AE, Elghamrawy S. Detection of Coronavirus (COVID-19) Associated Pneumonia based on Generative Adversarial Networks and a Fine-Tuned Deep Transfer Learning Model using Chest X-ray Dataset. *arXiv preprint* arXiv. 2020;2004.01184.
13. Hammoudi K, Benhabiles H, Melkemi M, Dornaika F, Arganda-Carreras I, Collard D, Scherpereel A. Deep Learning on Chest X-ray Images to Detect and Evaluate Pneumonia Cases at the Era of COVID-19. *arXiv preprint* arXiv. 2020;2004.03399.
14. Hemdan EE, Shouman MA, Karar ME. Covidx-net: A framework of deep learning classifiers to diagnose covid-19 in x-ray images. *arXiv preprint arXiv*. 2020;2003.11055.
15. Effects Of Coronavirus Disease (COVID-19) On Your Body, https://www.weforum.org/agenda/2020/04/this-graphic-shows-what-covid-19-does-to-your-body/, Accessed 23 April 2020.
16. Chalmers J, Kaul A, Phillips B. The wisdom of crowds: Mutual fund investors' aggregate asset allocation decisions. *Journal of Banking & Finance*. 2013;37(9):3318-33.
17. Bassamboo A, Cui R, Moreno A. Wisdom of Crowds in Operations: Forecasting Using Prediction Markets. Available at SSRN 2679663, 2015 Oct 25.
18. Walter T, Back A. Crowdsourcing as a Business Model: An exploration of emergent textbooks harnessing the wisdom of crowds. *AIS.* 2010.
19. Maghdid HS, Asaad AT, Ghafoor KZ, Sadiq AS, Khan MK. Diagnosing COVID-19 pneumonia from X-ray and CT images using deep learning and transfer learning algorithms. *arXiv preprint* arXiv. 2020;2004.00038.
20. Afshar P, Heidarian S, Naderkhani F, Oikonomou A, Plataniotis KN, Mohammadi A. COVID-CAPS: A Capsule Network-based Framework for Identification of COVID-19 cases from X-ray Images. *arXiv preprint arXiv.* 2020;2004.02696.
21. Alqudah AM, Qazan S, Alquran H, Qasmieh IA, Alqudah A. COVID-2019 DETECTION USING X-RAY IMAGES AND ARTIFICIAL INTELLIGENCE HYBRID SYSTEMS. *ResearchGate*, 2020 March.
22. Zhang J, Xie Y, Li Y, Shen C, Xia Y. Covid-19 screening on chest x-ray images using deep learning based anomaly detection. *arXiv preprint arXiv.* 2020;2003.12338.
23. Nwankpa C, Ijomah W, Gachagan A, Marshall S. Activation functions: Comparison of trends in practice and research for deep learning. *arXiv preprint* arXiv. 2018;1811.03378.
24. Covid-19 Chest x-ray dataset, https://GitHub.com/ieee8023/covid-chestxray-dataset, Accessed 21 April 2020.
25. Chest x-ray images, https://www.kaggle.com/paultimothymooney/chest-xray-pneumonia, Accessed 21 April 2020
26. Kingma DP, Ba J. Adam: A method for stochastic optimization. *arXiv preprint* arXiv. 2014;1412.6980.
27. Rodríguez P, Bautista MA, Gonzalez J, Escalera S. Beyond one-hot encoding: Lower dimensional target embedding. *Image and Vision Computing.* 2018;75:21-31.
28. Cui Y, Jia M, Lin TY, Song Y, Belongie S. Class-balanced loss based on effective number of samples. In *Proceedings of the IEEE Conference on Computer Vision and Pattern Recognition,* 2019 (pp. 9268-9277).
29. Nadarajah S. A generalized normal distribution. *Journal of Applied Statistics*. 2005;32(7):685-94.
30. Acute Respiratory Distress Syndrome -Ventilator Support, https://foundation.chestnet.org/patient-education-resources/acute-respiratory-distress-syndrome-ards/, Accessed 23 April 2020.